\documentclass[a4paper,11pt]{article}
\usepackage{pos}
\usepackage[capitalize]{cleveref}
\usepackage{wrapfig}
\usepackage[utf8]{inputenc}
\usepackage{textgreek}

\title{Comparing Calculations of Seasonal Variations of Atmospheric Muons in Deep Underground Detectors}
\ShortTitle{Comparing Calculations of Seasonal Variations of Atmospheric Muons}

\author*[a]{Amanda Alves}
\author[a]{Lilly Pyras}
\author[a]{Dennis Soldin}
\author[b]{Stef Verpoest}

\affiliation[a]{Department of Physics and Astronomy, University of Utah, USA}
\affiliation[b]{Bartol Research Institute, Department of Physics and Astronomy, University of Delaware, USA}

\emailAdd{amanda.alves@utah.edu}
\emailAdd{lilly.pyras@utah.edu}
\emailAdd{dennis.soldin@utah.edu}
\emailAdd{verpoest@udel.edu}

\abstract{
Cosmic rays interact with nuclei in the Earth's atmosphere to produce extensive air showers, which give rise to the atmospheric muon flux. Temperature fluctuations in the atmosphere influence the rate of muons measured in deep underground experiments. This contribution presents predictions of the daily muon flux at a depth of 2000 m.w.e., calculated using MUTE, a software tool which combines MCEq, a numerical solver of the matrix cascade equations in the atmosphere, with PROPOSAL, a propagation code for leptons in matter. The flux estimates are obtained assuming different cosmic-ray flux and hadronic interaction models. The results are compared to previous approaches, based on different methods, to calculate seasonal variations of atmospheric muons in deep underground detectors.
}

\ConferenceLogo{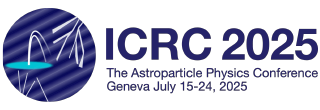}

\FullConference{39th International Cosmic Ray Conference (ICRC2025)\\
 15–24 July 2025\\
Geneva, Switzerland\\}

\begin{document}
\maketitle

\section{Introduction} 

High-energy cosmic rays initiate extensive air showers (EASs) in the upper atmosphere which produce a large flux of atmospheric leptons at the ground. This includes muons which are predominantly produced from decays of charged pions and kaons during the EAS development. In the energy range where the interaction lengths of pions and kaons are comparable to their decay lengths, higher temperatures lead to lower density, and subsequently to more decays, resulting in higher muon production rates. Thus, the muon flux at ground exhibits seasonal variations which are correlated with the atmospheric temperature. While the correlation is small for low energies, where most mesons decay, at energies above several TeV the muon flux is fully correlated with the temperature. There is a long history of measurements of seasonal variations of the atmospheric muon flux in deep-underground detectors and for the interpretation of these measurements theoretical predictions of the expected muon fluxes are crucial. 

This paper builds upon the formalism described in Ref.~\cite{Verpoest:2024dmc}, where different approaches for the calculation of seasonal variations of atmospheric muons are discussed in great detail. In particular, in \cref{sec:muon_fluxes} we will present updated calculations of muon energy spectra and production depths using the MCEq (Matrix Cascade Equations) code~\cite{Fedynitch:2015zma} and discuss their dependence on the hadronic interaction model used to describe the EAS development, as well as the impact of the cosmic-ray flux assumption. In \cref{sec:daily_rates} we will introduce improvements in the description of the muon propagation through dense media using the software package MUTE (MUon inTensity codE)~\cite{Fedynitch:2021ima} and derive the expected daily muon rates in underground detectors. The results will be compared to previous calculations based on different approaches.  Finally, we will discuss the implications of these improvements for measurements of atmospheric muons in \cref{sec:conclusions}.

\section{Muon Flux Calculations}
\label{sec:muon_fluxes}

The evolution of particles in an air shower is described by the coupled cascade equations~\cite{Gaisser:2016uoy,Verpoest:2024dmc}
\begin{align}
\label{eq:cascade_eq}
\frac{d\phi_i(E, X)}{\mathrm{d}X} = 
& -\frac{\phi_i(E, X)}{\lambda_{\text{int},i}(E)} 
  - \frac{\phi_i(E, X)}{\lambda_{\text{dec},i}(E, X)} \nonumber \\
& + \sum_j \int_{E}^{\infty} \mathrm{d}E_j \, 
    \frac{\mathrm{d}n_{j(i)\rightarrow i}(E)}{\mathrm{d}E} 
    \frac{\phi_j(E_j, X)}{\lambda_{\text{int},j}(E_j)} \nonumber \\
& + \sum_j \int_{E}^{\infty} \mathrm{d}E_j \, 
    \frac{\mathrm{d}n_{j(i)\rightarrow i}^{\text{dec}}(E)}{\mathrm{d}E} 
    \frac{\phi_j(E_j, X)}{\lambda_{\text{dec},j}(E_j, X)}\; ,
\end{align}
where $\phi_i(E, X)\, \mathrm{d}E$ is the flux of type $i$ particles with energies between $E$ and $E+dE$ at atmospheric slant depth $X$. The first two terms in \cref{eq:cascade_eq} are loss terms which describe the interaction and decay of particles $i$, depending on the interaction and decay lengths, $\lambda_\text{int}$ and $\lambda_\text{dec}$, respectively. The final two terms, where $\mathrm{d}n/\mathrm{d}E$ are the inclusive particle production spectra, are source terms for the generation of particle type $i$ as a result of the interaction and decay of particles of type $j$. The slant depth $X$ in \cref{eq:cascade_eq} at an observation height $h_0$ in the atmosphere is determined along the trajectory $l$ of the EAS central core by
\begin{equation}
    X(h_0, \theta) = \int_{h_0}^{\infty} \mathrm{d}l \, \rho_{\text{air}}(h(l, \theta))\;,
    \label{eq:slant_depth}
\end{equation}
where $\theta$ is the zenith angle, and the mass density of air, $\rho_\text{air}$, is a function of atmospheric height $h(l,\theta)$.  Particle fluxes in air showers are sensitive to changes in atmospheric temperature because density and temperature are directly correlated, assuming an ideal gas. The muon production spectrum, differential in muon energy $E_\mu$ and slant depth X, is given by
\begin{equation}
    \label{eq:production_spectrum}
    P(E_{\mu}, \theta, X) = \frac{\mathrm{d}\phi_{\mu}(E_{\mu}, \theta, X)}{\mathrm{d}X}\;.
\end{equation}
The integral over the production spectrum yields the flux of muons at the surface, differential in energy:
\begin{equation}
    \label{eq:phi}
    \phi_\mu(E_\mu, \theta) = \int_0^{X_0} \text{d}X P(E_\mu, \theta, X)\;.
\end{equation}
The muon production spectrum depends on the temperature $T(X)$ at slant depth $X$ because of its relationship to the atmospheric density profile (\cref{eq:slant_depth}). In a detector with effective area $A_\text{eff}$, the rate of muons with energy $E_{\mu}$ from a zenith angle direction $\theta$ is determined by
\begin{equation}
    \label{eq:rate1}
    R(\theta) = \int \mathrm{d}X \int_{E_{\mu}^{\text{min}}}^{\infty} \mathrm{d}E_{\mu} \, A_{\text{eff}}(E_{\mu}, \theta) \, P(E_{\mu}, \theta, X)\;.
\end{equation}
For a compact detector at a depth that is large compared to its vertical size, the effective area is the projected physical area of the detector at the zenith angle $\theta$ averaged over the azimuth angle, $A_\mathrm{proj}(\theta)$, multiplied by an energy-dependent efficiency factor. Assuming each muon reaching the detector is detected, this factor is equal to the survival probability $\epsilon(E_\mu, L(\theta))$ of muons with energy $E_\mu$ and track length $L(\theta)$ in the surrounding medium. The survival probability is determined by energy losses in the medium, mainly due to ionization and pair production. The expected muon rate in the detector can then be written as
\begin{align}
    \label{eq:rate2}
    R(\theta) &= A_\mathrm{proj}(\theta) \int \mathrm{d}X \int \mathrm{d}E_{\mu} \, \epsilon(E_\mu, L(\theta)) \; P(E_{\mu}, \theta, X)  \\
    \label{eq:rate21}
    &= A_\mathrm{proj}(\theta) \; \int \mathrm{d}X \, P_\mathrm{int, \epsilon}(\theta, X)\\
    \label{eq:rate22}
    &= A_\mathrm{proj}(\theta) \; I_\mathrm{\epsilon}(\theta)\;,
\end{align}
where $I_\mathrm{\epsilon}(\theta)$ is the flux of muons able to reach the detector for a given $\theta$, obtained by integrating over the efficiency-weighted production profile $P_\mathrm{int, \epsilon}(\theta, X)$. The survival probability can be approximated by a simple step function
\begin{equation}
\label{eq:step_function}
\epsilon(E_\mu, L(\theta)) = 
\begin{cases} 
1 \;, \;E_\mu\geq E_\mu^\text{min} \\
0 \;, \;E_\mu < E_\mu^\text{min},
\end{cases}
\end{equation}
where $E_\mu^\text{min}$ is the (average) minimum energy required to reach the detector at a distance $L(\theta)$ without being absorbed within the medium. However, more sophisticated approaches to account for the propagation of muons in a dense medium can be considered to determine $\epsilon(E_\mu, L(\theta))$, which will be discussed in \cref{sec:daily_rates}. Integration of \cref{eq:rate2} over the solid angle $\Omega$ yields the total expected muon rate in the detector,
\begin{equation}
    \label{eq:rate3}
    R = \int R(\theta) \, \mathrm{d}\Omega\;.
\end{equation}

However, due to the inherent complexity of air showers, the cascade equations in \cref{eq:cascade_eq} can generally not be solved analytically and calculations of the production profiles of muons, \emph{i.e.}, $P(E_{\mu}, \theta, X)$ in \cref{eq:production_spectrum}, typically rely on semi-analytical approximations, parameterizations based on Monte-Carlo simulations, or complex numerical methods. In the following, we will use MCEq to obtain numerical solutions of the cascade equations and compare the resulting muon distributions with those from analytical approximations and parameterization-based approaches.

\subsection{Muon Energy Spectra} 
\label{sec:energy_spectra}

\begin{figure}[tb]
    \centering
    \vspace{-0.5em}
    \includegraphics[width=\linewidth]{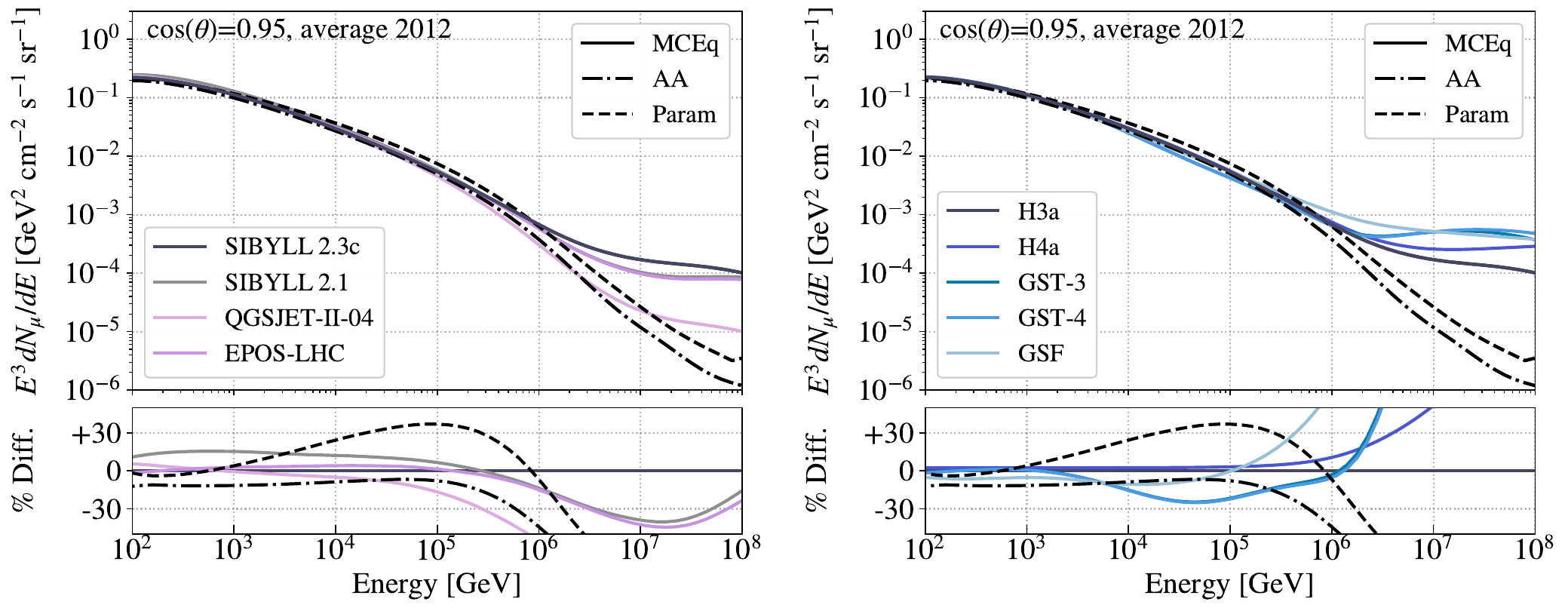}
    \vspace{-1.8em}
    
    \caption{Muon energy spectra (scaled with $E^3$) obtained from MCEq based on different hadronic models (left) and cosmic-ray flux assumptions (right). Also shown for comparison are predictions based on analytical approximations (AA) and on a parameterizations of Monte-Carlo simulations (Param). Shown in the bottom panels are the model differences with respect to MCEq predictions using H3a and SIBYLL\,2.3c.}
    \label{fig:muon_spectrum}
    \vspace{-0.5em}
\end{figure}

\Cref{fig:muon_spectrum} shows the muon energy spectra obtained from MCEq (scaled with $E^3$), assuming SIBYLL\,2.1~\cite{Ahn:2009wx}, SIBYLL\,2.3c~\cite{Fedynitch:2018cbl}, EPOS-LHC~\cite{Pierog:2013ria,Pierog:2015ifw}, and QGSJet-II-04~\cite{Ostapchenko:2005nj,Ostapchenko:2013pia} as the hadronic interaction model (left), and for the cosmic-ray flux models H3a/H4a~\cite{Gaisser:2011klf}, GST-3/GST-4~\cite{Gaisser:2013bla}, and GSF~\cite{Dembinski:2017zsh} (right). The figure also shows the percent differences for the different hadronic and flux model predictions with respect to SIBYLL\,2.3c and H3a. The muon spectra are averaged over the year 2012 based on daily atmospheric data at the South Pole from the Atmospheric Infrared Sounder (AIRS) on board NASA's Aqua satellite~\cite{AIRS2013}. The AIRS pressure and temperature data are used to determine the atmospheric density profile, $\rho_\mathrm{air}$, for each day of the year, assuming an ideal gas law. For comparison, the figure also includes predictions based on analytical approximations~(AA) of the cascade equations \cite{Gaisser:2016uoy} and calculations based on parameterizations of simulations~(Param) \cite{Gaisser:2021cqh}.

The MCEq predictions based on different hadronic and flux model assumptions show differences on the order of up to around $\pm 20\%$ and $\pm 15\%$, respectively, up to a few TeV. While the AA predictions agree with MCEq on a 10\% level up to energies of around 100\,TeV, in agreement with studies presented in Ref.~\cite{Gaisser:2019xlw}, the parameterization based approach yields large differences above a few TeV. At higher muon energies significant differences are observed. This is due to the modeling of prompt muon production via the decay of heavy hadrons which differs significantly between the models and is not included at all in AA and Param calculations. However, due to the steep energy spectrum of muons, the total muon rate recorded in underground detectors is highly dominated by muons close to the detector energy threshold and thus differences at high energies are negligible. 

\subsection{Muon Production Profiles} 
\label{sec:production_depth}

\begin{wrapfigure}{r}{0.55\textwidth}
    \centering
    \vspace{-1.4em}
    \includegraphics[width=1.\linewidth]{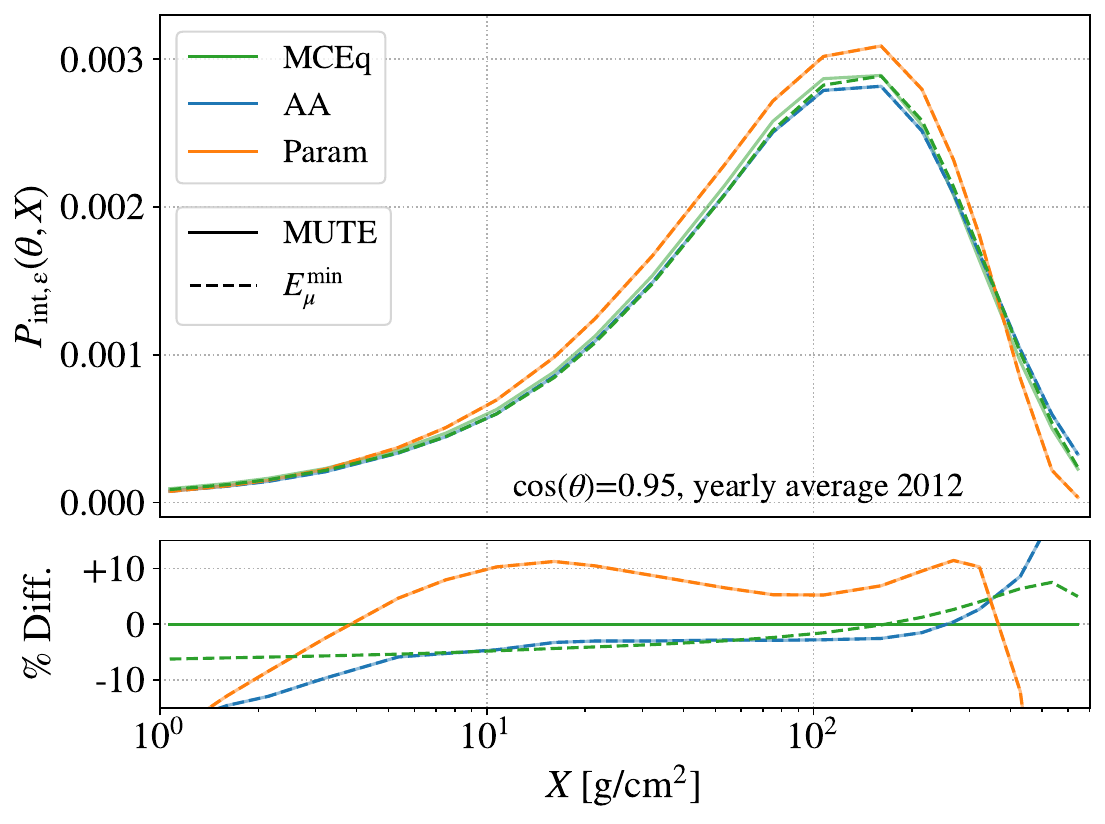}
    \vspace{-1.8em}
    
    \caption{{Yearly average efficiency-weighted muon production profiles (normalized) for $\cos(\theta) = 0.95$, obtained using the MCEq, AA, and Param approaches (see text for details). The profiles are shown assuming a simple energy threshold of $E_\mu > 660$\,GeV (dashed lines) and survival probabilities obtained from MUTE (solid lines). Percent differences with respect to MCEq (MUTE) are shown in the bottom panel.}}
    \label{fig:muon_prod_depth}
\end{wrapfigure}

\Cref{fig:muon_prod_depth} shows the (normalized) efficiency-weighted muon production profiles, $P_\mathrm{int, \epsilon}(\theta, X)$ in \cref{eq:rate22}, obtained from MCEq, analytical approximations, and the parameterization approach. The profiles are shown for a zenith angle direction of $\cos(\theta)=0.95$, as an example, averaged over the year 2012 based on AIRS data. Shown are the production profiles for muons with energies above $E_\mu^\mathrm{min}=660$\,GeV (dashed lines), corresponding to the minimum energy $E_\mu^\mathrm{min}$ required for a muon to reach a detector at a depth of 2000\,m.w.e, assuming energy losses due to ionization according to the approximation in Ref.~\cite{Gaisser:2016uoy}. Also shown are the equivalent distributions using more realistic survival probabilities of muons, obtained from MUTE (solid lines), as described further below in \cref{sec:daily_rates}. While the treatment of muon propagation ($E_\mu^\mathrm{min}$ vs. MUTE) does not affect the shape of the profiles significantly, the differences of calculations based on MCEq, AA, and Param are clearly visible. They are approximately constant between depths of 10\,g/cm$^2$ and a few 100\,g/cm$^2$, where most muons are produced, however, above around 400\,g/cm$^2$ the  calculations show significant differences.

\section{Daily Underground Muon Rates}
\label{sec:daily_rates}

In order to calculate the daily muon rates in underground detectors the probability of survival for muons to reach the detector needs to be taken into account, as discussed in \cref{sec:muon_fluxes}. This can be done, for example, using a simple energy threshold $E_\mu^\mathrm{min}$ determined from average ionization energy losses of muons. However, MUTE provides the possibility to estimate survival probabilities based on simulated muon propagation using the software code PROPOSAL (PRopagator with Optimal Precision and Optimized Speed for All Leptons)~\cite{koehne2013proposal,Dunsch:2018nsc}.

\begin{figure}[tb]
    \vspace{-1em}
    \mbox{\hspace{-1.2em}
    \includegraphics[width=0.53\linewidth]{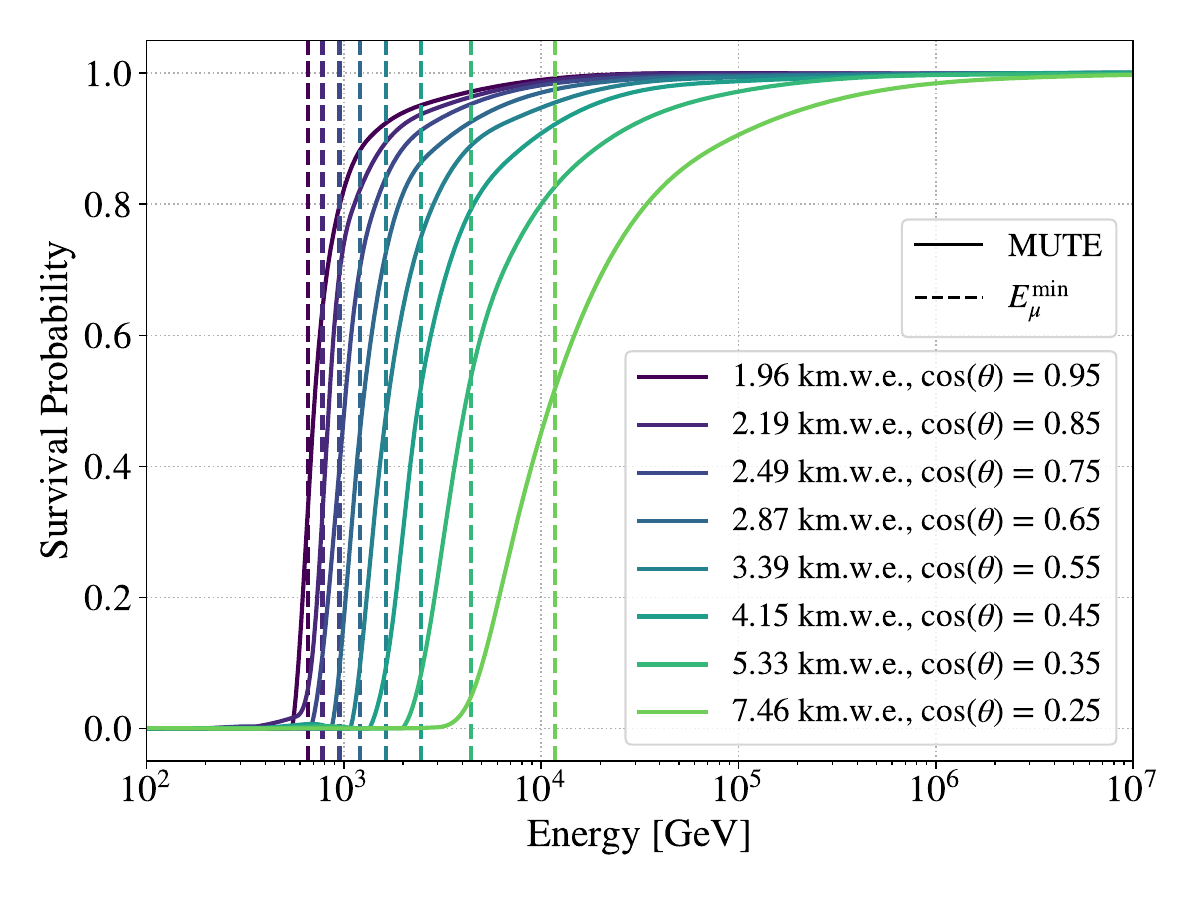}
    \hspace{-1.2em}
    \includegraphics[width=0.53\linewidth]{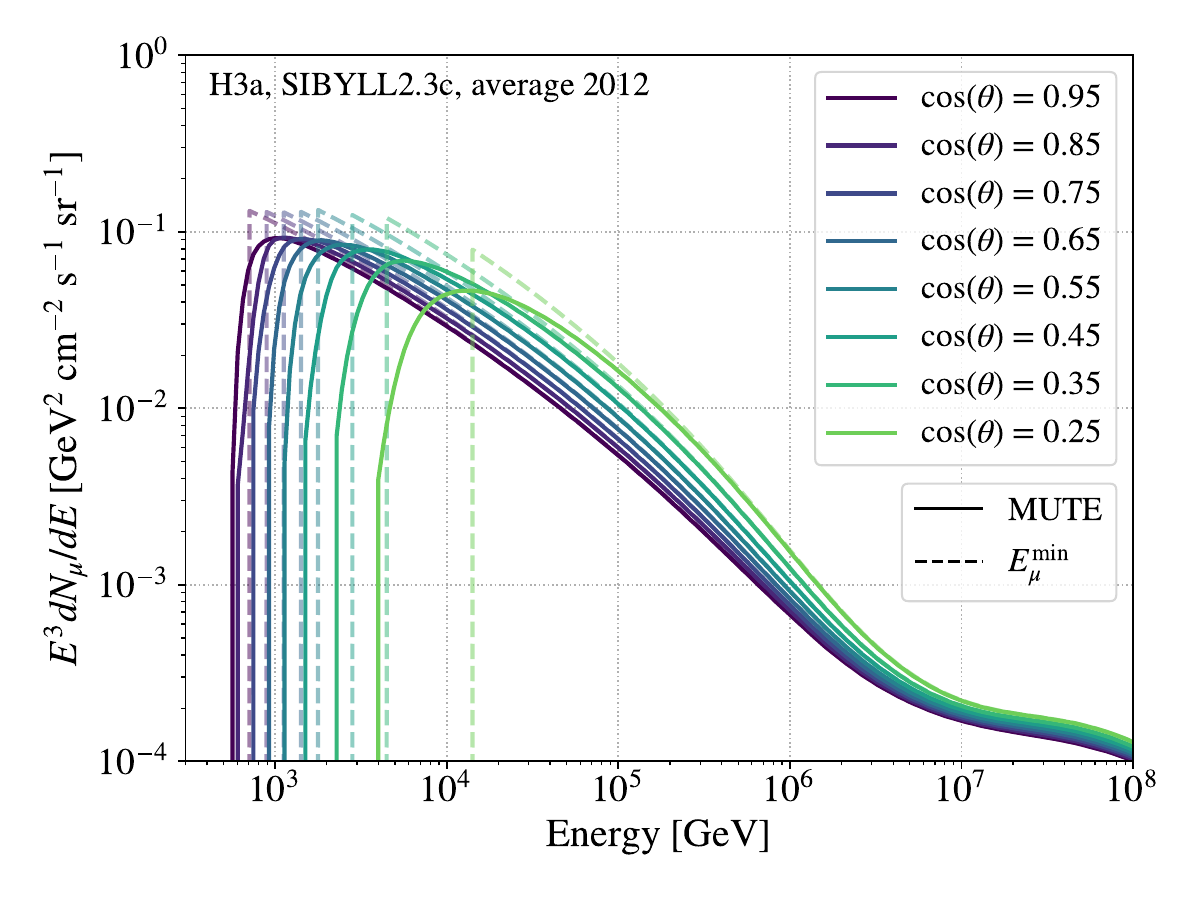}
    }
    \vspace{-2.2em}
    
    \caption{Survival probabilities as a function of muon energy and for different zenith angle directions (left), and corresponding muon energy spectra obtained from MCEq after applying the survival probabilities (right). Two approaches are used, a simple energy threshold $E_\mu^\mathrm{min}$, as well as an approach based on propagating muons in MUTE, as described in the text.}
    \label{fig:efficiencies}
    \vspace{-0.5em}
\end{figure}

\Cref{fig:efficiencies} (left) shows the survival probabilities obtained from propagating 10,000 muons with PROPOSAL to a detector at a vertical depth of 2000\,m.w.e. using the MUTE code for 10 zenith angle directions, equidistant in $\cos(\theta)$, and 120 muon energies, distributed uniformly in $\log_{10}(E_\mu)$ over the entire muon energy range. Also shown is the equivalent track length of muons through the medium in terms of water-equivalent depth. For comparison, the survival probabilities based on the simple step function in \cref{eq:step_function} are also shown, assuming muon energy thresholds $E_\mu^\mathrm{min}$ for different zenith angle directions, as described in Ref.~\cite{Verpoest:2024dmc}. \Cref{fig:efficiencies} (right) shows the average energy spectra for the year 2012 and for different zenith angle directions, obtained from MCEq, after applying the MUTE survival probabilities shown in the left figure.

\begin{figure}[b]
    \vspace{-1.em}
    \centering
    \includegraphics[width=1.01\linewidth]{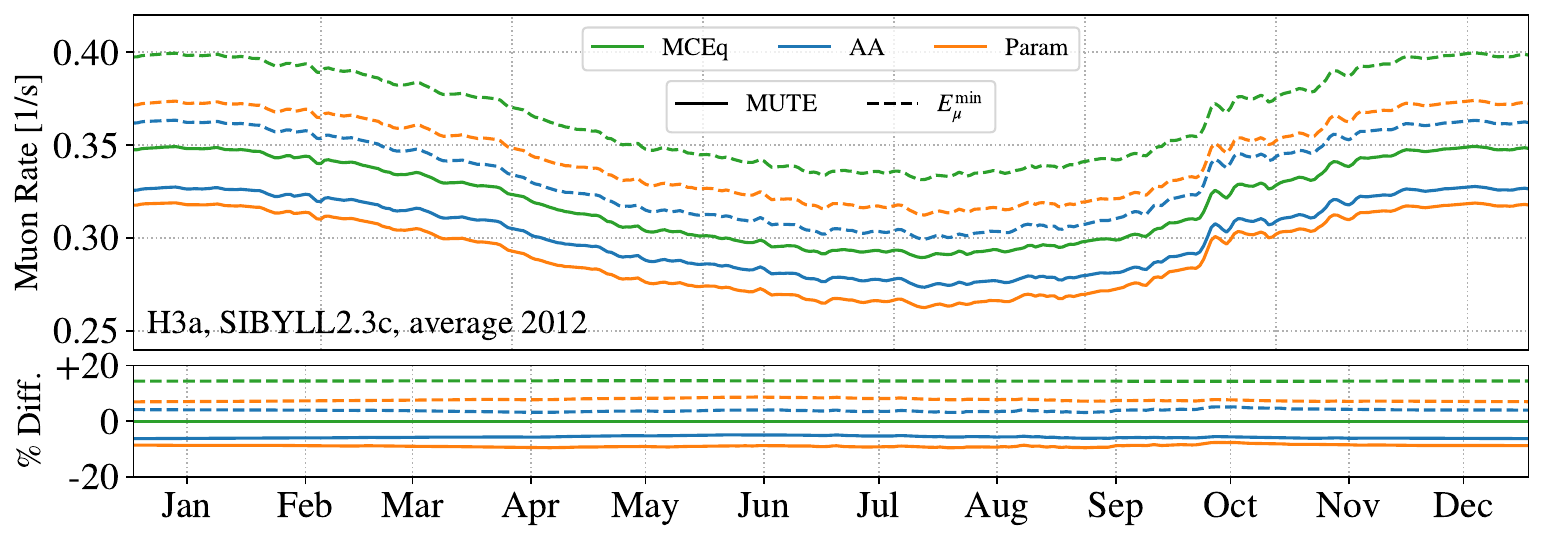}
    \caption{Daily rates for a cylindrical detector with a radius of 5\,m and height of 20\,m at a depth of 2000\,m.w.e. The rates are calculated using the MCEq, AA, and Param methods, and two different approaches to account for muon propagation in a dense medium (MUTE and $E_\mu^\mathrm{min}$). See text for details.}
    \label{fig:daily_rates}
    \vspace{-1.em}
\end{figure}

For illustration purposes, the daily expected rates are calculated according to \cref{eq:rate3}, using AIRS daily atmospheric temperature profiles, for a hypothetical cylindrical detector with a radius of 5\,m and height of 20\,m at a depth of 2000\,m.w.e. The resulting event rates are shown in \cref{fig:daily_rates} based on production profiles obtained from MCEq, the analytical approximations, and the parametrization approach. They are compared for a simple energy threshold $E_\mu^\mathrm{min}$ (dashed lines) and the more sophisticated approach using survival probabilities obtained from MUTE (solid lines). Also shown are the percent differences between the predictions with respect to MCEq assuming MUTE survival probabilities (bottom panel).

While the daily variations overall show a remarkable agreement between the different approaches, using the MUTE survival probabilities yields around 15\% lower daily muon event rates than the simple minimum energy thresholds, as expected from the comparison of survival probabilities and spectra in \cref{fig:efficiencies}. In addition, significant differences between the overall rates can be observed between the AA, Param, and in particular, MCEq approaches, which is expected due to the differences of the energy spectra in \cref{fig:muon_spectrum} at a few TeV.

\begin{wrapfigure}{r}{0.55\textwidth}
    \centering
    \includegraphics[width=1.\linewidth]{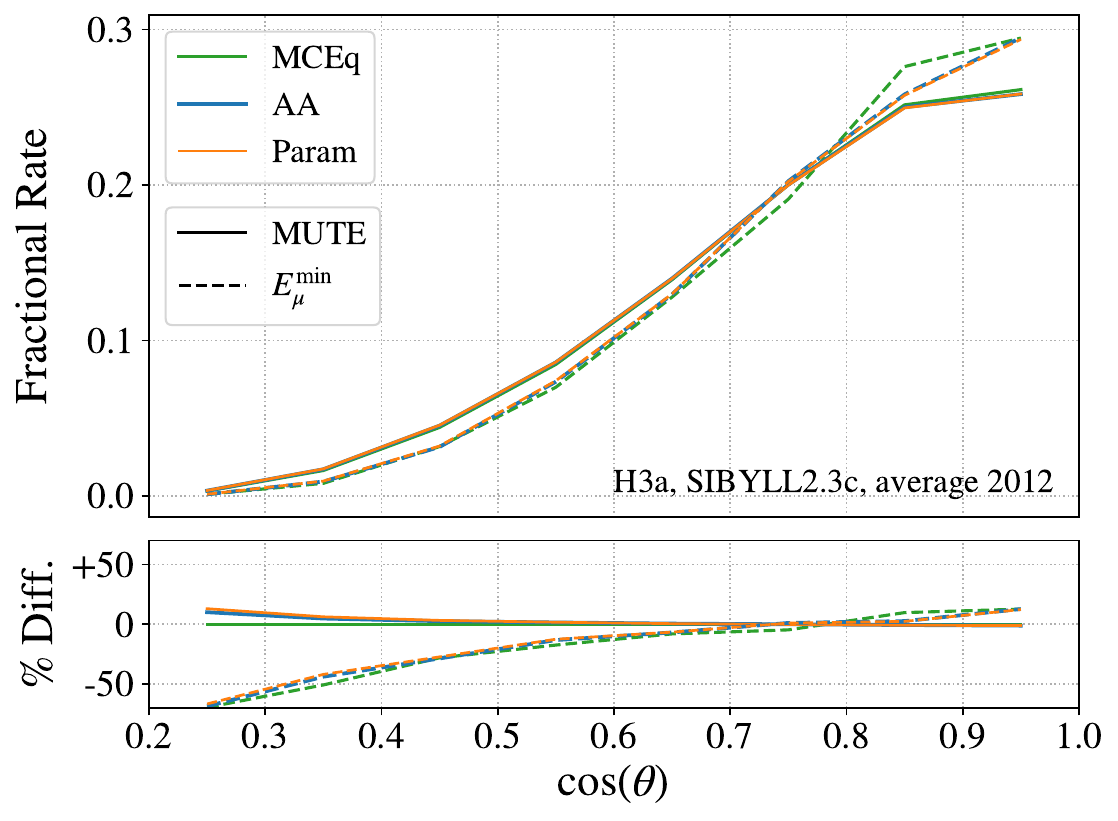}
    \caption{Fractional rates of muons as a function of the zenith angle, calculated for a detector assumed in \cref{fig:daily_rates} using the MCEq, AA, and Param methods (see text for details), and two different approaches to account for muon propagation in a dense medium (MUTE and $E_\mu^\mathrm{min}$).}
    \label{fig:zenith_dist}
\end{wrapfigure}

\Cref{fig:zenith_dist} shows the corresponding fractional rates of the total yearly flux in dependence of the zenith angle direction. While the differences in the relative angular distribution between MCEq, AA, and Param are small, significant differences are observed for the different models of muon propagation, \emph{i.e.}, MUTE and $E_\mu^\mathrm{min}$, especially towards large zenith angles ($\cos(\theta)\rightarrow 0$) where the muon propagation becomes important due to the large path length in the medium. However, differences of up to 20\% can also be observed for very vertical directions ($\cos(\theta)> 0.9$) which may be due to radiative energy losses of high-energy muons that are not accounted for in the simple approach accounting for ionization only.

\section{Conclusions}
\label{sec:conclusions}

We have presented updated predictions of seasonal variations of atmospheric muon fluxes in deep underground detectors based on the computational tools MCEq and MUTE, following the formalism described in Ref.~\cite{Verpoest:2024dmc}. These tools include the treatment of hadronic interactions in the atmosphere, cosmic-ray fluxes, and muon propagation through dense media. The specific model assumptions introduce uncertainties in the spectra of muons of up to 20\% in the energy range relevant for underground detectors. A comparison with previous calculations based on analytical approximations of the cascade equations and parameterizations of Monte-Carlo distributions show differences of around 15\%. Moreover, predictions based on muon propagation estimated with MUTE yield 15\% lower rates in comparison to calculations using a simple threshold energy $E_\mu^\mathrm{min}$.

Preliminary measurements of the seasonal variations of muons measured with the IceCube Neutrino Observatory have shown discrepancies in the total rate with respect to calculations based on analytical approximations~\cite{Tilav:2019xmf}. Thus, it is of large interest to compare the updated calculations to data from IceCube or KM3Net. However, the calculations presented in this work are based on the inclusive muon fluxes and do not account for multiple muons from the same EAS arriving simultaneously at the detector. In order to compare calculations with data, the muon multiplicity needs to be taken into account as described in Ref.~\cite{Verpoest:2024dmc}. Together with the improvements discussed here, such calculations may be used in the future to study muon production in EASs. It has been shown, for example, that seasonal variations of muons are sensitive to the hadron composition in EASs and may be used to estimate the pion-to-kaon ratio~\cite{Desiati:2010wt}, or the contributions from heavy hadron (prompt) decays. Thereby, such studies can contribute to improve the understanding of multi-particle production in air showers.

\newpage
\section*{Acknowledgements}

We acknowledge the support and resources from the Center for High Performance Computing at the University of Utah. This work was supported by Travel \& Small Grants from the Office of Undergraduate Research at the University of Utah awarded to A.\,A. and by a Postdoc Travel Assistance Award from the Postdoc Affairs Office at the University of Utah awarded to L.\,P. S.\,V. acknowledges funding from the National Science Foundation (NSF) grant \#2209483.

%%%%%% REFERENCES...  %%%%%%

\bibliographystyle{ICRC}
\bibliography{references}

\end{document}